\documentclass[toc]{PoS}
\usepackage{amssymb,fontenc, times, mathptmx,graphicx}
\title{Verification of Monte Carlo transport codes against measured small
angle p-, d-, and t-emission in carbon fragmentation at 600 MeV/nucleon
}
\ShortTitle{p,d and t from carbon fragmentation}
\author{B. M. Abramov, P. N. Alexeev, Yu. A. Borodin, S. A. Bulychjov,
I. A. Dukhovskoy, 
A.~P. Krutenkova,
\speaker{V. V. Kulikov}, M. A. Martemianov, M. A. Matsyuk, E. N. Turdakina, 
A. I. Khanov \\
Institute for Theoretical and Experimental Physics, NRC 'Kurchatov Institute',
Moscow, Russia\\
E-mail: \email{kulikov@itep.ru}}
\author{S. G. Mashnik\\
Los Alamos National Laboratory (LANL), Los Alamos, NM 87545, USA}

\abstract{Momentum spectra of hydrogen isotopes have been measured
at 3.5$^\circ$ from  $^{12}$C fragmentation on a Be target. Momentum
spectra cover both the region of fragmentation maximum and the
cumulative region. Differential cross sections span five orders of
magnitude. The data are compared
 to predictions of four Monte Carlo codes: QMD, LAQGSM, BC, and INCL++. There are
 large differences between the data and predictions of some models in 
the high momentum region.
The INCL++ code gives the best and almost perfect description of the data.}

\FullConference{XXII International Baldin Seminar on High Energy Physics
Problems \\
                 September 15-20, 2014\\
                 JINR, Dubna, Russia}

\begin{document}

\section{Introduction}
The study of nucleus-nucleus interactions is one of the main aims of
modern nuclear physics. During the last few years, apart from an
investigation of fundamental properties of these interactions,
special attention has been paid to precise phenomenological
descriptions of the processes used in applications such as heavy ion
therapy, radiation shielding, and radioactive ion beam design. Along
the way a few simulation programs for nucleus-nucleus interactions
have been created. They demand an experimental verification as well
as refinement of their basic approaches. One of the aims of the
FRAGM experiment performed at the TWA (Tera Watt Accumulator) heavy ion
facility at ITEP was to obtain  high precision data on nuclear
fragmentation in the energy range accessible at this accelerator. In
the framework of this experiment, the data has been taken for carbon
ion fragmentation on a Beryllium target  both in a wide incident energy range
from 0.2 to 3.2 GeV/nucleon and in a wide energy range of the
fragments from hydrogen isotopes to isotopes of the projectile
nucleus. The measurements were performed in the projectile
fragmentation region, i.e. in the so-called inverse kinematics. This
method has substantial advantages over measurements in the target
fragmentation region. At first, for the fragments moving in the forward
direction, a relativistic boost provides larger acceptance in the
projectile rest frame in case of an equal acceptance in the
laboratory frame. Secondly, in inverse kinematics, there are no
problems with a detection of fragments that are at rest in a
projectile rest frame, because they are moving in a laboratory frame
with a speed of the projectile nucleus. 
Yields of cumulative (high energy) protons were analyzed
\cite{Abramov2013}  in a framework of a multiquark cluster model and
compared to a few models of ion-ion interactions in
\cite{Abramov2015}. In this publication we give preliminary results
on heavier hydrogen isotope emissions at 600 MeV incident carbon ion
energy and compare them with the predictions of four 
Monte Carlo
codes.

\section{Experiment}
The experiment was carried out at the  TWA heavy-ion complex
 at ITEP  which includes an ion laser source, a linac,
 a booster,  and an accelerator-accumulator  ring. Ions of 200-1000
MeV/A  could  be accumulated in this ring for successive use in
experiments on  high-energy-density physics or accelerated to a
maximal energy of 4 GeV/nucleon. During our measurements, each four seconds
the carbon ions C$^{+4}$ were accelerated in  the booster up to
300 MeV/nucleon.   Then, while injection to the accelerator-accumulator
ring they were totally stripped, captured in the ring and accelerated up to 600 MeV/nucleon.
After that the beam was steered to
the internal target of 50 $\mu$m Be foil strip providing the spill.
It made simultaneously possible to have both a high luminosity 
because of
multiple passage of the ions through the target and a small size of
the source needed for a high momentum resolution of the subsequent magnetic analysis.
 The products of the carbon nucleus fragmentation
outgoing at 3.5$^\circ$ were captured by the double-focus beam line 
(42 meters long). Sets of few a scintillation counters were placed at
intermediate and final focuses  for multiple
measurements of dE/dx and time of flight (TOF).  Fragments with different
charge and mass were unambiguously selected on two-dimensional plots
dE/dx vs TOF.  The set up has been described  in more detail in
\cite{Abramov2013}. The fragment
momentum spectra were obtained by beam line energy scan  in steps of 50-200 MeV/c
and fragments were selected using the  procedure mentioned  above. 
As a monitor, we used a telescope of three scintillation counters that viewed
 the target at 2$^\circ$.

\section{Models}
Recently, intermediate-energy ions have been used
in various fields of nuclear physics and applications.
This supports the demand both for deepening our knowledge of
 fundamental properties of ion-ion interactions and for developing
methods of precise simulation of these processes on up-to-date
levels. Both 
ways need large amount of various experimental
data for testing new theoretical ideas and application programs.
Here we use new data of the FRAGM experiment for verification of
four ``event-generators''
widely used transport codes. They are LAQGSM03.03
\cite{LAQGSM}, QMD \cite{QMD}, BC \cite{BC} and INCL++ \cite{INCL}.
LAQGSM03.03 (Los Alamos version of the Quark Gluon String Model)  is
supported and updated by LANL in the USA. It is a main part of the MCNP6
transport code  \cite{MCNP6}. The other three codes, QMD (Quantum
Molecular Dynamics), BC (Binary Cascade), and INCL++(C++ (5.1.14)
version of the Liege Intranuclear Cascade model) are free access
programs from the GEANT4 package \cite{GEANT4} supported by CERN. We
used the version Geant4 10.0 (released 6 December 2013). 
In general,
all of these
codes consider ion-ion interactions as a sequence of the
same processes such as intranuclear cascade, formation of excited
prefragments and their successive deexcitation by evaporation
(preceded by preequilibrium emission, in the case of LAQGSM03.03), 
Fermi breakup, and fission. But, an actual realization of these steps are
different in different models. A description of these differences
are far beyond the scope of this publication. Short and useful
information on this subject can be found in the
GEANT4 Physics Reference manual \cite{GEANT4} and in Ref. \cite{LAQGSM} (for LAQGSM03.03).

\begin{figure}
\begin{center}
\includegraphics[width=1.0\textwidth]{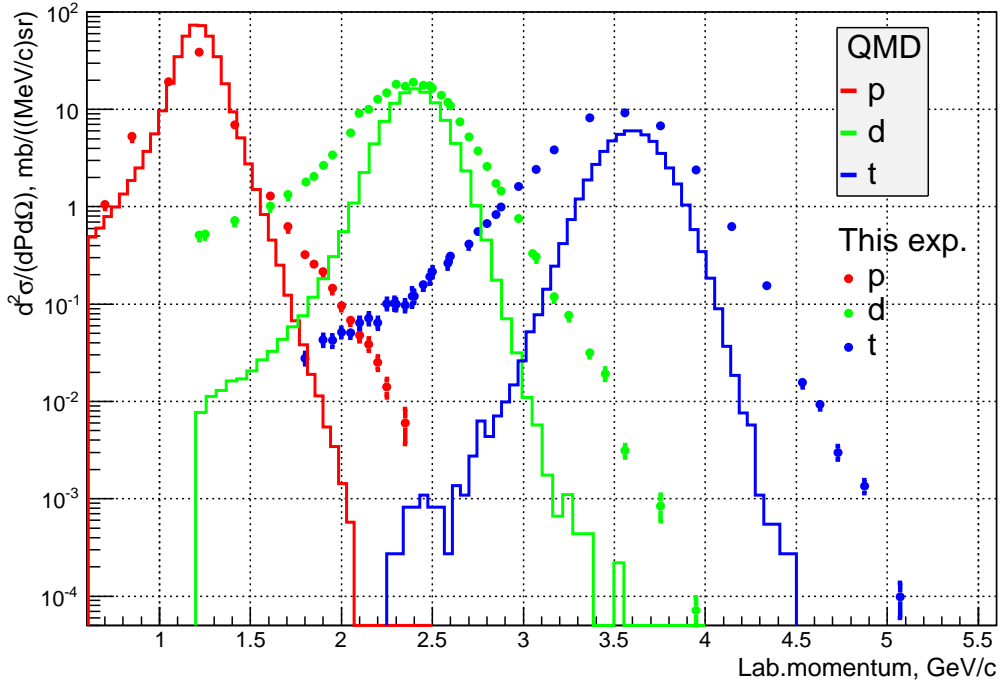}
 \caption{ Laboratory momentum spectra of protons (full red circles at the left), 
deuterons (full green circles at the center), and tritons (full blue circles at the 
right) emitted at the 3.5$^\circ$ from fragmentation of 600 MeV/nucleon carbon ions  on 
Be target in comparison with the predictions of the QMD model (red, green, and 
blue histograms, correspondingly).}
 \label{fig:QMD}
\end{center}
\end{figure}

\begin{figure}
\begin{center}
\includegraphics[width=1.\textwidth]{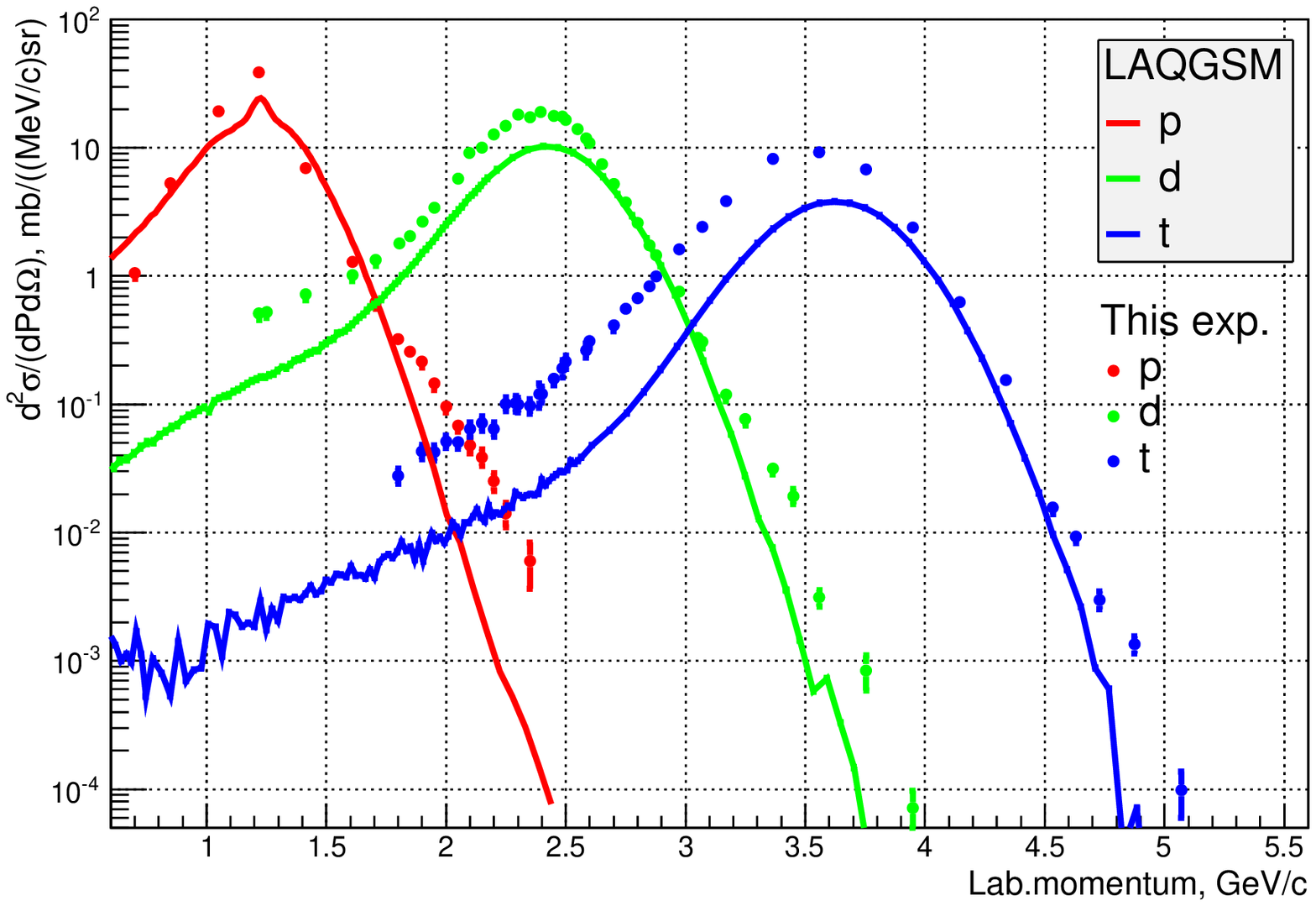}
 \caption{
 The same experimental data as in Fig.1 but in comparison with the predictions 
of the LAQGS03.03 model.}
\label{fig:LA}
\end{center}
\end{figure}

\begin{figure}
\begin{center}
\includegraphics[width=1.\textwidth]{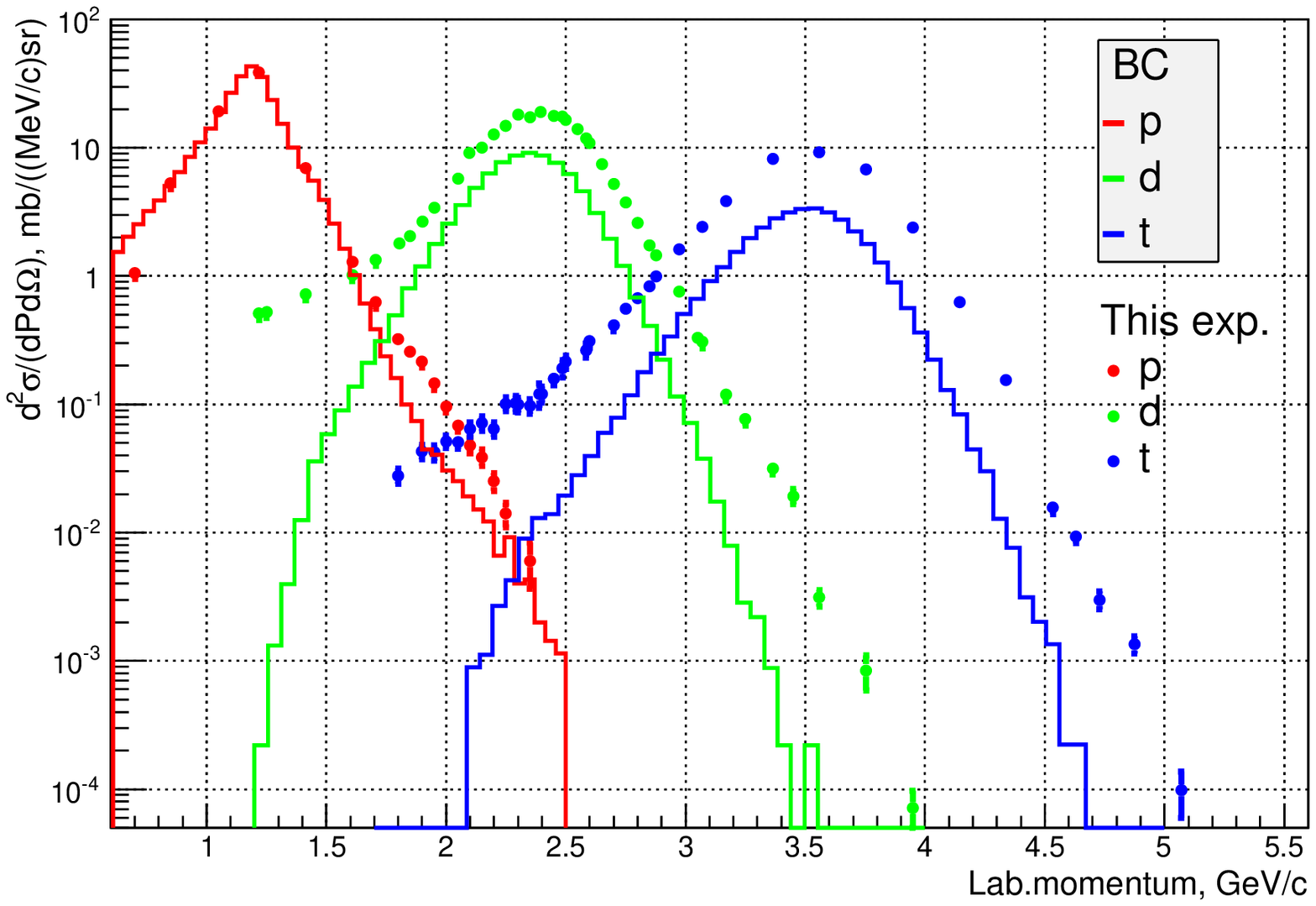}
 \caption{
 The same experimental data as in Fig.1 but in comparison with the predictions 
of the BC model.}
\label{fig:BC}
\end{center}
\end{figure}

\begin{figure}
\begin{center}
\includegraphics[width=1.\textwidth]{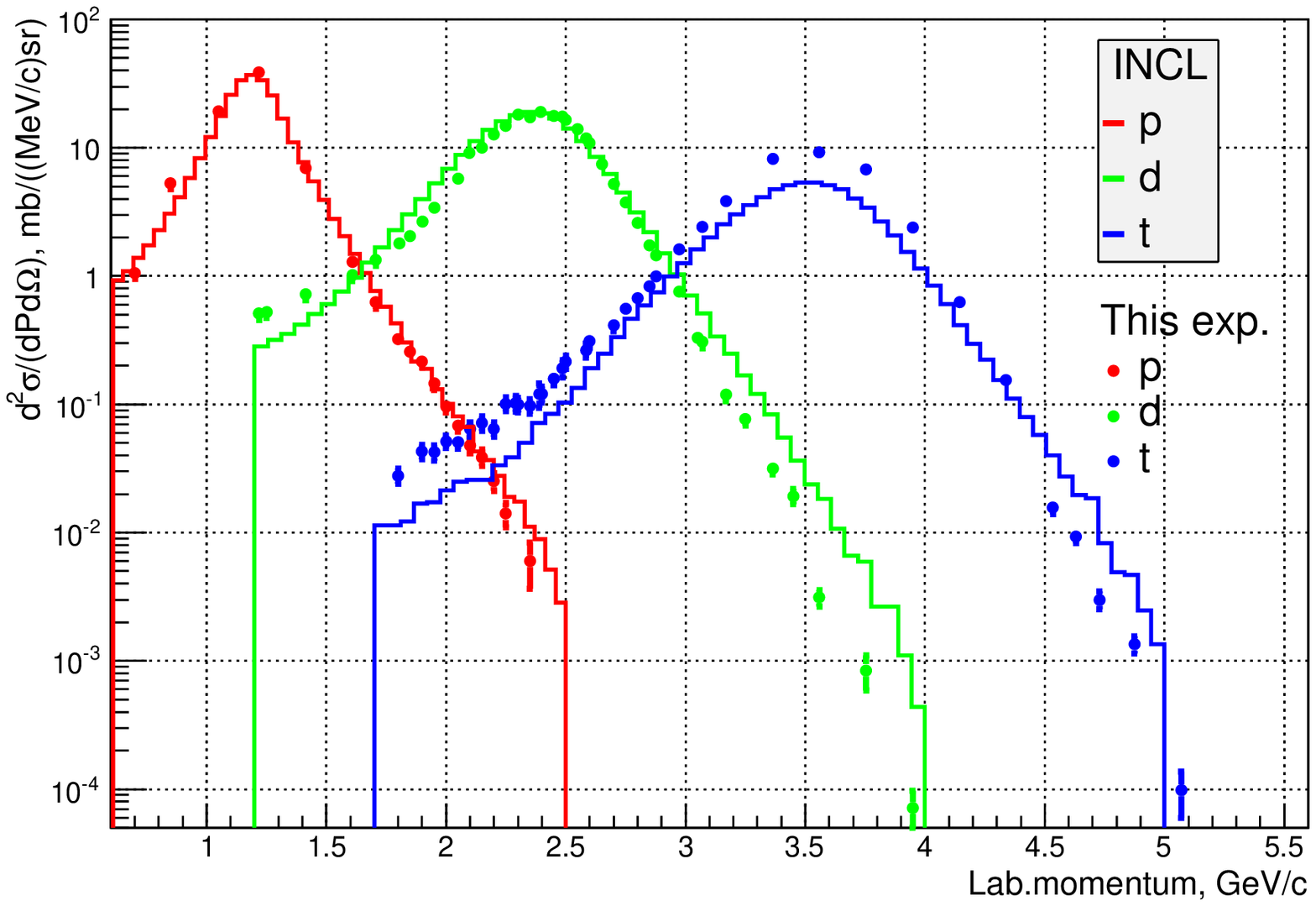}
 \caption{
 The same experimental data as in Fig.1 but in comparison with the predictions 
of the INCL++ model.}
\label{fig:INCL}
\end{center}
\end{figure}

\section{Comparison of model predictions with experimental data}

Differential cross sections $d^2\sigma/dpd\Omega$ in a laboratory frame for proton, 
deuteron and triton yields
at an angle  of 3.5$^\circ$ and results of the simulations,
using the models mentioned above,
are
given in 
Fig. \ref{fig:QMD} --- \ref{fig:INCL} 
in logarithmic scale as a function of
 laboratory momentum of fragments.
 For normalization of calculations, the total inelastic cross section of 
$^{12}$C + $^9$Be interactions equal to 823.8 mb was used in accordance with the 
LAQGSM03.03 prescription.
The data of the FRAGM experiment were normalized to the BC predictions at the
fragmentation peak maximum for protons. The BC was chosen  because the shape 
of the fragmentation peak is in good agreement with that observed in the experiment. 
It can be seen from Fig. \ref{fig:INCL} that this normalization is of the same 
quality for the INCL++ model.
 For each fragment the spectrum demonstrates the 
so called fragmentation peak 
with maximum at the fragment velocity approximately  equal to that of the projectile 
carbon ion. For protons the peak maximum is at $\sim$1.2 GeV/c, for deuterons -- 
at $\sim$2.4 GeV/c, and for tritons -- at $\sim$3.6 GeV/c. The experimental points  
go down at higher momentum (in the so called cumulative region) by 5 orders of 
differential cross section magnitude and show a less steep fall to lower momentum 
(midrapidity) region. As expected, all models give a reasonable  qualitative 
description of the data. But there are substantial quantitative differences 
between the data and model predictions as well as between the different models.

The QMD model (see  Fig. \ref{fig:QMD}) gives good predictions for differential 
cross sections at fragmentation maxima, but the widths of the peaks for all 
fragments are too narrow. To give quantitative results we fitted these peaks 
of longitudinal momentum distributions with gaussians near their  maxima  in the
projectile rest frame. For this experiment the r.m.s of the peaks are 72$\pm$5, 
~134$\pm$3, and 160$\pm$4 MeV/c for protons, deuterons, and tritons, respectively. 
While for QMD, these values are 56, 88, and 100 MeV/c. Here and later the errors 
for  model calculations will not be given because they always can be made much 
smaller than experimental ones. The data of our experiment are in reasonable 
agreement with existing measurement \cite{Greiner} 63$\pm$4,~113$\pm$11, and 
162$\pm$14 MeV/c. The narrow widths of the fragmentation peaks predicted by QMD 
result partly in a large underestimation of the differential cross section in the
cumulative and midrapidity regions. The differences exceed two orders of magnitude 
at the edges of studied momentum intervals. For LAQGSM03.03 (see  Fig. \ref{fig:LA}), 
the widths of the fragmentation peaks for deuterons and tritons are in reasonable 
agreement with the data. They are 147 and 159 MeV/c in projectile rest frame, 
respectively. For protons the predicted shape of the peak is not gaussian. 
Unfortunately the momentum step of our data are too large to check this prediction. 
LAQGSM03.03 underestimates the differential cross section at the fragmentation 
peak maxima by a factor of 2 and does not reproduce the cumulative tail of 
proton spectra at the 2.0-2.5 GeV/c laboratory momentum region, but shapes of 
deuteron and triton momentum spectra in the midrapidity region are reproduced well.

  The BC model (see  Fig. \ref{fig:BC}) gives slightly smaller values for 
fragmentation peak widths than LAQGSM03.03. The r.m.s. valuers are 62, 122, and 
140 MeV/c for protons, deuterons, and tritons in projectile rest frame. The BC 
underestimates differential cross sections in the cumulative and midrapidity regions 
for deuterons and tritons, but gives reasonable description of the proton 
cumulative tail.
 Fig. \ref{fig:INCL} with the results of calculations by the INCL++ model demonstrates 
almost ideal agreement with the data of our experiment. The r.m.s. of the 
fragmentation peaks in projectile rest frame equal to 66, ~140, and 167 MeV/c 
(for protons, deuterons, and tritons) are very close to our measurements.  
The proton spectrum is reproduced with high precision  even in the cumulative 
region. The midrapidity region is also well described. The differential cross 
sections in the cumulative regions for deuterons and tritons are even slightly 
overestimated while all other models underestimate them in  varying degrees.

\section{Conclusion}
The comparison performed shows a large potential of modern transport codes 
to describe new high precision data on fragmentation in ion-ion interactions. 
Apart from the QMD model, all other tested models   (LAQGSM03.03, BC, and INCL++) 
give reasonable descriptions of proton, deuteron, and triton momentum spectra 
at 3.5$^\circ$ from 600 MeV/nucleon carbon fragmentation on Be targets. 
Some problems arise for the cumulative region where BC and LAQGSM03.03 
substantially underestimate the differential cross section. In an absence of a
recognized theory of cumulative particle production, the phenomenological 
mechanism of excited prefragment productions and their subsequent Fermi breakup  
is widely used in the transport codes. Good agreement between the data and 
predictions of the INCL++ model shows that this approach is very successful 
for a phenomenological description of the data of our experiment on cumulative 
proton, deuteron, and triton emission in carbon fragmentation.

The authors are grateful to operating personnel of ITEP TWA and to technicians 
of the FRAGM experiment for a large contribution in carrying out the measurements.  
We thank Dr. Roger L. Martz for a very careful 
reading of the manuscript and useful suggestions on its improvement.
Part of the work, performed at LANL,  was carried out under 
DOE Contract No. DE-AC52-06NA25396. This work has been supported 
by the Russian Foundation for Basic Research (grant RFBR 15-02-06308).

\end{document}